\documentclass[10pt]{article}
\usepackage{latexsym,graphicx}
\newcommand{\be}{\begin{equation}}
\newcommand{\ee}{\end{equation}}
\def\n{\noindent}
\catcode `\@=11
\catcode `\@=12
\begin{document}
\begin{center}
\large{\bf {Inhomogeneous Bulk Viscous Fluid Universe with Electromagnetic Field and Variable 
$\Lambda$-Term}} \\
\vspace{10mm}
\normalsize{Anirudh Pradhan \footnote{Corresponding author}, Vandana Rai $^2$, Kanti Jotania$^3$}\\
\vspace{5mm}
\normalsize{$^{1}$ Department of Mathematics, Hindu Post-graduate College, 
Zamania-232 331, Ghazipur, India \\
E-mail : pradhan@iucaa.ernet.in} \\
\normalsize{$^{2}$Department of Mathematics, Post-graduate College, Ghazipur-233 001, India \\
E-mail : vandana\_rai005@yahoo.co.in}\\
\vspace{5mm}
\normalsize{$^3$Department of Physics, Faculty of Science, The M. S. University of Baroda, 
Vadodara-390 002, India \\
E-mail : kanti@iucaa.ernet.in}\\ 
\end{center}
\vspace{10mm}
\begin{abstract} 
Cylindrically symmetric inhomogeneous cosmological model for bulk viscous fluid distribution with 
electromagnetic field is obtained. The source of the magnetic field is due to an electric current 
produced along the z-axis. $F_{12}$ is the non-vanishing component of electromagnetic field tensor. 
To get the deterministic solution, it has been assumed that the expansion $\theta$ in the model is 
proportional to the shear $\sigma$. The values of cosmological constant for these models are found 
to be small and positive at late time which are consistent with the results from recent supernovae 
Ia observations. Physical and geometric aspects of the models are also discussed in presence and 
absence of magnetic field.
\end{abstract}
\smallskip
\n Keywords : cosmology, variable cosmological term, electromagnetic field, inhomogeneous universe\\
\n PACS number: 98.80.Jk, 98.80.-k
\section{Introduction}
Inhomogeneous cosmological models play an important role in understanding some essential 
features of the universe such as the formation of galaxies during the early stages of 
evolution and process of homogenization. The early attempts at the construction of such 
models have been done by Tolman \cite{ref1} and Bondi \cite{ref2} who considered spherically 
symmetric models. Inhomogeneous plane-symmetric models were considered by 
Taub \cite{ref3,ref4} and later by Tomimura \cite{ref5}, Szekeres \cite{ref6}, Collins 
and Szafron \cite{ref7}, Szafron and Collins \cite{ref8}. Recently, Senovilla 
\cite{ref9} obtained a new class of exact solutions of Einstein's equations without 
big bang singularity, representing a cylindrically symmetric, inhomogeneous cosmological 
model filled with perfect fluid which is smooth and regular everywhere satisfying 
energy and causality conditions. Later, Ruiz and Senovilla \cite{ref10} have examined 
a fairly large class of singularity free models through a comprehensive study of 
general cylindrically symmetric metric with separable function of $r$ and $t$ as metric 
coefficients. Dadhich et al. \cite{ref11} have established a link between the FRW 
model and the singularity free family by deducing the latter through a natural and simple
in-homogenization and anisotropization of the former. Also, Patel et al. 
\cite{ref12} presented a general class of inhomogeneous cosmological models filled with 
non-thermalized perfect fluid by assuming that the background space-time admits two 
space-like commuting Killing vectors and has separable metric coefficients. Singh, Mehta and 
Gupta \cite{ref13} obtained inhomogeneous cosmological models of perfect fluid distribution 
with electro-magnetic field. Recently, Pradhan et al. \cite{ref14} have investigated plane-symmetric 
inhomogeneous cosmological models in various contexts. Cylindrically symmetric space-time play 
an important role in the study of the universe on a scale in which anisotropy and inhomogeneity 
are not ignored. Roy and Singh \cite{ref15}, Bali and Tyagi \cite{ref16,ref17}, 
Chakrabarty et al. \cite{ref18} and Pradhan et al. \cite{ref19} have investigated cylindrically 
symmetric inhomogeneous cosmological models in presence of electromagnetic field.
\par
The occurrence of magnetic field on galactic scale is well-established fact today, and their 
importance for a variety of astrophysical phenomena is generally acknowledged as pointed
out by Zeldovich et al. \cite{ref20}. Also Harrison \cite{ref21} has suggested that magnetic field 
could have a cosmological origin. As a natural consequences, we should include magnetic fields in 
the energy-momentum tensor of the early universe. The choice of anisotropic cosmological models in
Einstein system of field equations leads to the cosmological models more general than Robertson-Walker 
model \cite{ref22}. The presence of primordial magnetic field in the early stages of the evolution 
of the universe has been discussed by several authors \cite{ref23}$-$\cite{ref32}. Strong magnetic 
field can be created due to adiabatic compression in clusters of galaxies. Large-scale magnetic field 
gives rise to anisotropies in the universe. The  anisotropic pressure created by the magnetic fields 
dominates the evolution of the shear anisotropy and it decays slower than the case when the pressure 
was isotropic \cite{ref33,ref34}. Such fields can be generated at the end of an inflationary epoch 
\cite{ref35}$-$\cite{ref39}. Anisotropic magnetic field models have significant contribution in the 
evolution of galaxies and stellar objects. Bali and Ali \cite{ref40} obtained a magnetized cylindrically 
symmetric universe with an electrically neutral perfect fluid as the source of matter. Chakrabarty et al. 
\cite{ref18} and Pradhan et al. \cite{ref41} have investigated magnetized viscous fluid cosmological 
models in various contexts.
\par
There are significant observational evidence for the detection of Einstein's cosmological constant, 
$\Lambda$ or a component of material content of the universe that varies slowly with time and space 
to act like $\Lambda$. Some of the recent discussions on the cosmological constant ``problem'' and on 
cosmology with a time-varying cosmological constant by Ratra and Peebles \cite{ref42}, and Sahni and 
Starobinsky \cite{ref43} point out that in the absence of any interaction with matter or radiation, 
the cosmological constant remains a ``constant''. However, in the presence of interactions with matter 
or radiation, a solution of Einstein equations and the assumed equation of covariant conservation 
of stress-energy with a time-varying $\Lambda$ can be found. This entails that energy has to be conserved 
by a decrease in the energy density of the vacuum component followed by a corresponding increase in the 
energy density of matter or radiation (see also Carroll, Press and Turner \cite{ref44}, Peebles \cite{ref45}, 
Padmanabhan \cite{ref46}). There is a plethora of astrophysical evidence today, from supernovae measurements 
(Perlmutter et al. \cite{ref47}, Riess et al. \cite{ref48}, Garnavich et al. \cite{ref49}, Schmidt et al. 
\cite{ref50}, Blakeslee et al. \cite{ref51}, Astier et al. \cite{ref52}), the spectrum of fluctuations in 
the Cosmic Microwave Background (CMB) \cite{ref53}, baryon oscillations \cite{ref54} and other astrophysical 
data, indicating that the expansion of the universe is currently accelerating. The energy budget of the 
universe seems to be dominated at the present epoch by a mysterious dark energy component, but the precise 
nature of this energy is still unknown. Many theoretical models provide possible explanations for the dark 
energy, ranging from a cosmological term \cite{ref55} to super-horizon perturbations \cite{ref56} and 
time-varying quintessence scenarios \cite{ref57}. These recent observations strongly favour a significant 
and a positive value of $\Lambda$ with magnitude $\Lambda(G\hbar/c^{3}) \approx 10^{-123}$. 
In Ref. \cite{ref48}, Riess et al. have recently presented an analysis of 156 SNe including a few at $z > 1.3$ 
from the Hubble Space Telescope (HST) ``GOOD ACS'' Treasury survey. They conclude to the 
evidence for present acceleration $q_{0} < 0$ $(q_{0} \approx -0.7)$. Observations (Knop et al. \cite{ref58}; 
Riess et al., \cite{ref48}) of Type Ia Supernovae (SNe) allow us to probe the expansion history of the 
universe leading to the conclusion that the expansion of the universe is accelerating.\\
\par
Most studies in cosmology involve a perfect fluid. Large entropy 
per baryon and the remarkable degree of isotropy of the cosmic microwave background 
radiation, suggest that we should analyze dissipative effects in cosmology. Further,
there are several processes which are expected to give rise to viscous effect. 
These are the decoupling of neutrinos during the radiation era and the recombination 
era \cite{ref59}, decay of massive super string modes into massless modes \cite{ref60},
gravitational string production \cite{ref61,ref62} and particle creation effect in 
grand unification era \cite{ref63}. It is known that the introduction of bulk 
viscosity can avoid the big bang singularity. Thus, we should consider the presence 
of a material distribution other than a perfect fluid to have realistic 
cosmological models (see Gr\o n \cite{ref64} for a review on cosmological 
models with bulk viscosity). A uniform cosmological model filled with fluid which possesses
pressure and second (bulk) viscosity was developed by Murphy \cite{ref65}. The solutions that
he found exhibit an interesting feature that the big bang type singularity appears in the 
infinite past. 
\par
Recently, Pradhan et al. \cite{ref66} have obtained inhomogeneous perfect fluid universe with 
electromagnetic field. Motivated by the situation discussed above, in this paper, we have 
obtained a new cylindrically symmetric inhomogeneous cosmological model for bulk viscous 
fluid distribution in presence and absence of electromagnetic field. The coefficient of 
bulk viscosity is assumed to be a power function of mass density. This work generalize 
the previous work of Pradhan et al. \cite{ref66}. Some physical and geometric behaviour of 
the models in presence and absence of magnetic field are also discussed. This paper is organized 
as follows. The metric and the field equations are laid down in Section 2. In Section 3, we deal 
with the solutions of the field equations in presence of bulk viscous fluid with electromagnetic 
field and variable cosmological term. We have also described the physical and geometric aspects of 
the models. In Section 4, we obtain the solutions of the field equations in absence of the magnetic 
field. Finally in Section 5 concluding remarks are given.   
\section{The Metric and Field  Equations}
We consider the metric in the form 
\begin{equation}
\label{eq1}
ds^{2} = A^{2}(dx^{2} - dt^{2}) + B^{2} dy^{2} + C^{2} dz^{2},
\end{equation}
where $A$ is the function of $t$ alone and  $B$ and $C$ are functions of $x$ and $t$.
The energy momentum tensor is taken as has the form 
\begin{equation}
\label{eq2}
T^{j}_{i} = (\rho + \bar{p})u_{i} u^{j} + \bar{p} g^{j}_{i} +  E^{j}_{i},
\end{equation}
where $E^{j}_{i}$ is the electromagnetic field given by Lichnerowicz \cite{ref67} 
\begin{equation}
\label{eq3}
E^{j}_{i} = \bar{\mu}\left[h_{l}h^{l}\left(u_{i}u^{j} + \frac{1}{2}g^{j}_{i}\right) 
- h_{i}h^{j}\right],
\end{equation}
and
\begin{equation}
\label{eq4}
\bar{p} = p - \xi u^{i}_{;i}.
\end{equation}
Here $\rho$, $p$, $\bar{p}$ and $\xi$ are the energy density, isotropic pressure, effective 
pressure and bulk viscous coefficient respectively and $u^{i}$ is the fluid four-velocity vector 
satisfying the condition 
\begin{equation}
\label{eq5}
g_{ij} u^{i}u^{j} = -1.
\end{equation}
$\bar{\mu}$ is the magnetic permeability and $h_{i}$, the magnetic flux vector 
defined by
\begin{equation}
\label{eq6}
h_{i} = \frac{1}{\bar{\mu}} \, {^*}F_{ji} u^{j},
\end{equation}
where the dual electromagnetic field tensor $^{*}F_{ij}$ is defined by Synge \cite{ref68} 
\begin{equation}
\label{eq7}
^{*}F_{ij} = \frac{\sqrt{-g}}{2} \epsilon_{ijkl} F^{kl}.
\end{equation}
Here $F_{ij}$ is the electromagnetic field tensor and $\epsilon_{ijkl}$ is the Levi-Civita 
tensor density.\\
The co-ordinates are considered to be comoving so that $u^{1}$ = $u^{2}$ = $u^{3}$ = $0$ and 
$u^{4} = \frac{1}{A}$. If we consider that the current flows  along the $z$-axis, then $F_{12}$ is the only 
non-vanishing component of $F_{ij}$. The Maxwell's equations  
\begin{equation}
\label{eq8}
F_{ij;k} + F_{jk;i}+ F_{ki;j}  = 0,
\end{equation}
\begin{equation}
\label{eq9}
\left[\frac{1}{\bar{\mu}}F^{ij}\right]_{;j} = 4 \pi J^{i}
\end{equation}
require that $F_{12}$ is the function of $x$-alone. We assume that the magnetic permeability is the 
functions of $x$ and $t$ both. Here the semicolon represents a covariant differentiation. \\

The Einstein's field equations (in gravitational units $c = 1, G = 1$) read 
\begin{equation}
\label{eq10}
R^{j}_{i} - \frac{1}{2} R g^{j}_{i}  + \Lambda g^{j}_{i} = - 8 \pi T^{j}_{i},
\end{equation}
for the line-element (\ref{eq1}) lead to the following system of equations:  
\[
\frac{1}{A^{2}}\left[- \frac{B_{44}}{B} - \frac{C_{44}}{C} + \frac{A_{4}}{A}\left(\frac{B_{4}}{B} + 
\frac{C_{4}}{C}\right) - \frac{B_{4} C_{4}}{B C} + \frac{B_{1}C_{1}}{BC}\right] - \Lambda 
\]
\begin{equation}
\label{eq11}
= 8 \pi \left(\bar{p} + \frac{F^{2}_{12}}{2\bar{\mu} A^{2} B^{2}} \right),
\end{equation}
\begin{equation}
\label{eq12}
\frac{1}{A^{2}}\left(\frac{A^{2}_{4}}{A^{2}}- \frac{A_{44}}{A} - \frac{C_{44}}{C} + 
\frac{C_{11}}{C} \right) - \Lambda =  8 \pi \left(\bar{p} + \frac{F^{2}_{12}}{2\bar{\mu} A^{2} B^{2}} \right), 
\end{equation}
\begin{equation}
\label{eq13}
\frac{1}{A^{2}}\left(\frac{A^{2}_{4}}{A^{2}} - \frac{A_{44}}{A} - \frac{B_{44}}{B} + 
\frac{B_{11}}{ B}\right) - \Lambda  =  8 \pi \left(\bar{p} - \frac{F^{2}_{12}}{2\bar{\mu} A^{2} B^{2}} \right), 
\end{equation}
\[
\frac{1}{A^{2}}\left[- \frac{B_{11}}{B} - \frac{C_{11}}{C} + \frac{A_{4}}{A}\left(\frac{B_{4}}{B} + 
\frac{C_{4}}{C}\right) - \frac{B_{1}C_{1}}{BC}  + \frac{B_{4} C_{4}}{B C}\right] + \Lambda
\]
\begin{equation}
\label{eq14}
= 8 \pi \left(\rho + \frac{F^{2}_{12}}{2\bar{\mu} A^{2} B^{2}}\right),
\end{equation}
\begin{equation}
\label{eq15}
\frac{B_{14}}{B} + \frac{C_{14}}{C} - \frac{A_{4}}{A}\left(\frac{B_{1}}{B} + \frac{C_{1}}{C}\right) = 0,
\end{equation}
where the sub indices $1$ and $4$ in A, B, C and elsewhere denote ordinary differentiation
with respect to $x$ and $t$ respectively.
\section{Solution of the Field Equations}
Equations (\ref{eq11})-(\ref{eq15}) are five independent equations in six unknowns $A$, $B$, $C$, 
$\rho$, $p$ and $F_{12}$. For the complete determinacy of the system, we need one extra condition. 
The research on exact solutions is based on some physically reasonable restrictions used to simplify 
the Einstein equations. \\
To get determinate solution we assume that the expansion $\theta$ in the model is proportional to the 
shear $\sigma$. This condition leads to 
\begin{equation}
\label{eq16}
A = \left(\frac{B}{C}\right)^{n},
\end{equation}
where $n$ is a constant. 
From Eqs. (\ref{eq11})-(\ref{eq13}), we have 
\begin{equation}
\label{eq17}
\frac{A_{44}}{A} - \frac{A^{2}_{4}}{A^{2}} + \frac{A_{4}B_{4}}{AB} + \frac{A_{4}C_{4}}{AC} - 
\frac{B_{44}}{B} - \frac{B_{4}C_{4}}{BC}  = \frac{C_{11}}{C} - \frac{B_{1}C_{1}}{BC} = \mbox{K (constant)}
\end{equation}
and
\begin{equation}
\label{eq18}
\frac{8\pi F^{2}_{12}}{\bar{\mu}B^{2}} = - \frac{C_{44}}{C} + \frac{C_{11}}{C} + \frac{B_{44}}{B} - 
\frac{B_{11}}{B}.
\end{equation}
We also assume that
\[
B = f(x)g(t)
\]
\begin{equation}
\label{eq19}
C = f(x)k(t).
\end{equation}
Using Eqs. (\ref{eq16}) and (\ref{eq17}) in (\ref{eq15}) and (\ref{eq17}) lead to
\begin{equation}
\label{eq20}
\frac{k_{4}}{k} = \frac{(2n - 1)}{(2n + 1)}\frac{g_{4}}{g},
\end{equation}
\begin{equation}
\label{eq21}
(n - 1)\frac{g_{44}}{g} - n\frac{k_{44}}{k} - \frac{g_{4}}{g}\frac{k_{4}}{k} = K,
\end{equation}
\begin{equation}
\label{eq22}
f f_{11} - f^{2}_{1} = Kf^{2}.
\end{equation}
Equation (\ref{eq20}) leads to
\begin{equation}
\label{eq23}
k = cg^{\alpha},
\end{equation}
where $\alpha = \frac{2n - 1}{2n + 1}$ and $c$ is the constant of integration. From Eqs. (\ref{eq21}) and 
(\ref{eq23}), we have
\begin{equation}
\label{eq24}
\frac{g_{44}}{g} + \beta\frac{g^{2}_{4}}{g^{2}} = N,
\end{equation}
where
$$
\beta = \frac{n\alpha(\alpha - 1) + \alpha}{n(\alpha - 1) + 1}, \, \, N = \frac{K}{n(1 - \alpha) - 1}. $$
Equation (\ref{eq22}) leads to 
\begin{equation}
\label{eq25}
f = \exp{\left(\frac{1}{2}K(x + x_{0})^{2}\right)},
\end{equation}
where $x_{0}$ is an integrating constant. Equation (\ref{eq24}) leads to
\begin{equation}
\label{eq26}
g = \left(c_{1}e^{bt} + c_{2}e^{-bt}\right)^{\frac{1}{(\beta + 1)}},
\end{equation}
where $b = \sqrt{(\beta + 1)N}$ and $c_{1}$, $c_{2}$ are integrating constants. Hence from (\ref{eq23}) and 
(\ref{eq26}), we have
\begin{equation}
\label{eq27}
k = c\left(c_{1}e^{bt} + c_{2}e^{-bt}\right)^{\frac{\alpha}{(\beta + 1)}}.
\end{equation}
Therefore we obtain
\begin{equation}
\label{eq28}
B = \exp{\left(\frac{1}{2}K(x + x_{0})^{2}\right)} \left(c_{1}e^{bt} + c_{2}e^{-bt}\right)^{\frac{1}{(\beta + 1)}},
\end{equation}
\begin{equation}
\label{eq29}
C = \exp{\left(\frac{1}{2}K(x + x_{0})^{2}\right)} c \left(c_{1}e^{bt} + c_{2}e^{-bt}\right)^{\frac{\alpha}{(\beta + 1)}},
\end{equation}
\begin{equation}
\label{eq30}
A = a \left(c_{1}e^{bt} + c_{2}e^{-bt}\right)^{\frac{n(1 - \alpha)}{(\beta + 1)}},
\end{equation}
where $a = \frac{c_{3}}{c}$, $c_{3}$ being a constant of integration. \\
After using suitable transformation of the co-ordinates, the model (\ref{eq1}) reduces to the form
\[
ds^{2} = a^{2}(c_{1}e^{bT} + c_{2}e^{-bT})^{\frac{2n(1 - \alpha)}{(\beta + 1)}}(dX^{2} - dT^{2}) 
+ e^{KX^{2}}(c_{1}e^{bT} + c_{2}e^{-bT})^{\frac{2}{(\beta + 1)}} dY^{2}
\]
\begin{equation}
\label{eq31}
+ e^{KX^{2}}(c_{1}e^{bT} + c_{2}e^{-bT})^{\frac{2\alpha}{(\beta + 1)}} dZ^{2},
\end{equation}
where $x + x_{0} = X$, $t = T$, $y = Y$, $cz = Z$.

The expressions for effective pressure $\bar{p}$ and density $\rho$  for the model (\ref{eq31}) are given by
\[
8\pi \bar{p} = \frac{1}{a^{2}(c_{1}e^{bT} + c_{2}e^{-bT})^{\frac{2n(1 - \alpha)}{(\beta + 1)}}} \Biggl[
\frac{b^{2}\{2n(1 - \alpha^{2}) + 2\beta + 2\alpha(\beta - \alpha)(1 - \alpha)\}}{2(\beta + 1)^{2}} \times
\]
\begin{equation}
\label{eq32}
\frac{(c_{1}e^{bT} - c_{2}e^{-bT})^{2}}{(c_{1}e^{bT} + c_{2}e^{-bT})^{2}} - \frac{b^{2}(3\alpha + 1)}
{2(\beta + 1)} + K^{2} X^{2}\Biggr] - \Lambda,
\end{equation}
\[
8 \pi \rho = \frac{1}{a^{2}(c_{1}e^{bT} + c_{2}e^{-bT})^{\frac{2n(1 - \alpha)}{(\beta + 1)}}} \Biggl[
\frac{b^{2}\{2n(1 - \alpha^{2}) + 2\alpha + (\beta - \alpha)(1 - \alpha)\}}{2(\beta + 1)^{2}} \times
\]
\begin{equation}
\label{eq33}
\frac{(c_{1}e^{bT} - c_{2}e^{-bT})^{2}}{(c_{1}e^{bT} + c_{2}e^{-bT})^{2}} - \frac{b^{2}(1 - \alpha)}
{2(\beta + 1)} - K (2 + 3 K X^{2}) \Biggr] + \Lambda.
\end{equation}
For the specification of $\xi$, we assume that the fluid obeys an equation of state of the form
\begin{equation}
\label{eq34} 
p = \gamma \rho,
\end{equation}
where $\gamma(0 \leq \gamma \leq 1)$ is a constant. Thus, given $\xi(t)$ we can solve for the cosmological 
parameters. In most of the investigation involving bulk viscosity it is assumed to be a simple power function 
of the energy density \cite{ref69}$-$\cite{ref73}  
\begin{equation}
\label{eq35} 
\xi(t) = \xi_{0} \rho^{m},
\end{equation}
where $\xi_{0}$ and $m$ are constants. For small density, $m$ may even be equal to unity as used in Murphy's 
work \cite{ref65} for simplicity. If $m = 1$, (\ref{eq35}) may correspond to a radiative fluid \cite{ref73}. 
Near a big bang, $0 \leq m \leq \frac{1}{2}$ is a more appropriate assumption \cite{ref74} to obtain realistic
 models. \\
\[
8\pi (p - \xi_{0} \rho^{m} \theta)  = \frac{1}{a^{2}(c_{1}e^{bT} + c_{2}e^{-bT})^{\frac{2n(1 - \alpha)}
{(\beta + 1)}}}\Biggl[\frac{b^{2}\{2n(1 - \alpha^{2}) + 2\beta + 2\alpha(\beta - \alpha)(1 - \alpha)\}}
{2(\beta + 1)^{2}}\times
\]
\begin{equation}
\label{eq36}
\frac{(c_{1}e^{bT} - c_{2}e^{-bT})^{2}}{(c_{1}e^{bT} + c_{2}e^{-bT})^{2}} - \frac{b^{2}
(3\alpha + 1)}{2(\beta + 1)} + K^{2} X^{2}\Biggr] - \Lambda,
\end{equation}
where $\theta$ is the scalar of expansion calculated for the flow vector $u^{i}$ and is given by
\begin{equation}
\label{eq37}
\theta = \frac{K_{2}}{(c_{1}e^{bT} + c_{2}e^{-bT})^{\frac{n(1 - \alpha)}
{(\beta + 1)}}}\frac{(c_{1}e^{bT} - c_{2}e^{-bT})}{(c_{1}e^{bT} + 
c_{2}e^{-bT})},
\end{equation}
where
\begin{equation}
\label{eq38}
K_{2} = \frac{b\{n(1 - \alpha) + (1 + \alpha)\}}{(\beta + 1)a}
\end{equation}
For simplicity and realistic models of physical importance, we consider the following two cases $(m = 0, 1)$.
On using (\ref{eq35}) in (\ref{eq32}), we obtain
\subsection{Model I: Solution when $m = 0$}
When $m = 0$, Eq. (\ref{eq35}) reduces to $\xi = \xi_{0}$. With the use of Eqs. (\ref{eq33}), (\ref{eq34}) 
and (\ref{eq37}), Eq.  (\ref{eq36}) reduces to
\[
8\pi (1 + \gamma)\rho = \frac{1}{a^{2}T_{1}^{\frac{2n(1 - \alpha)}{(1 + \beta)}}}\Biggl[\frac{b^{2}}{2(\beta + 1)^{2}}
\Big\{4n(1 - \alpha^{2}) + 2(\alpha + \beta) \,+  
\]
\begin{equation}
\label{eq39}
(\beta - \alpha)(1 - \alpha)(2\alpha + 1)\Big\}\left(\frac{T_{2}}{T_{1}}\right)^{2} - \, \frac{b^{2}(\alpha + 1)}
{(\beta + 1)} - 2K(1 + KX^{2})\Biggr] + \, \frac{8\pi \xi_{0}K_{2}}{T_{1}^{\frac{n(1 - \alpha)}{(1 + \beta)}}}
\left(\frac{T_{2}}{T_{1}}\right).  
\end{equation}
Eliminating $\rho(t)$ between (\ref{eq33}) and (\ref{eq39}), we get
\[
(1 + \gamma)\Lambda = \frac{1}{a^{2}T_{1}^{\frac{2n(1 - \alpha)}{(1 + \beta)}}}\Biggl[\frac{b^{2}}{2(\beta + 1)^{2}}
\Big\{2n(1 - \alpha^{2})(1 - \gamma) + 2(\beta - \alpha \gamma) \,+  
\]
\[
(\beta - \alpha)(1 - \alpha)(2\alpha - \gamma)\Big\}\left(\frac{T_{2}}{T_{1}}\right)^{2} + \frac{b^{2}\{
(\gamma - 1) - \alpha (\gamma + 3)\}}{2(\beta + 1)} 
\]
\begin{equation}
\label{eq40}
+ \, K\{KX^{2} + \gamma(2 + 3KX^{2})\}\Biggr] + \frac{8\pi \xi_{0}K_{2}}{T_{1}^{\frac{n(1 - \alpha)}{(1 + \beta)}}}
\left(\frac{T_{2}}{T_{1}}\right),  
\end{equation}
where
$$ T_{1} = (c_{1}e^{bT} + c_{2}e^{-bT}),$$
$$ T_{2} = (c_{1}e^{bT} - c_{2}e^{-bT}).$$
\begin{figure}[htbp]
\centering
\includegraphics[width=8cm,height=8cm,angle=0]{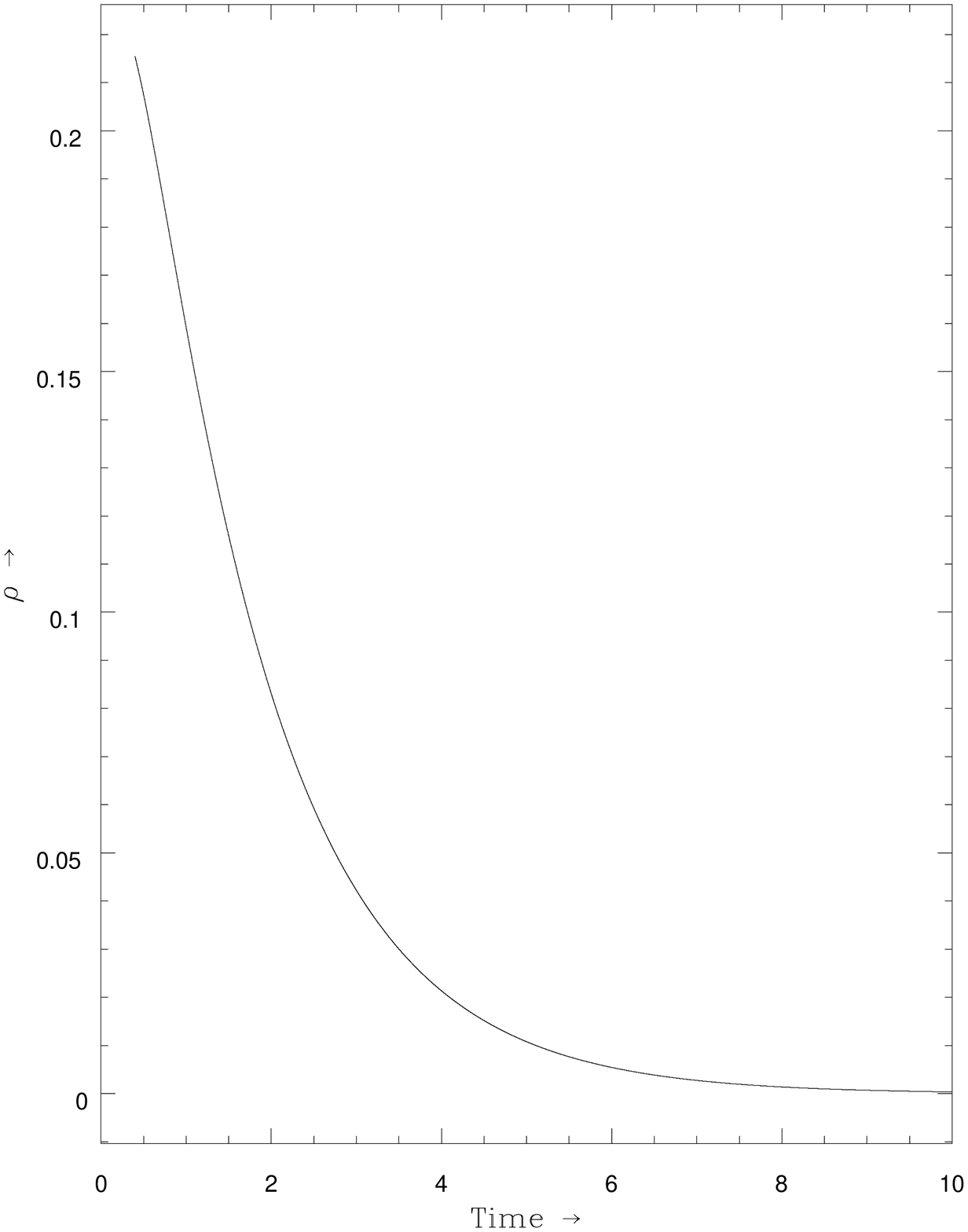}
\caption{The plot of energy density $\rho(T)$ Vs. T}
\end{figure}
\begin{figure}[htbp]
\centering
\includegraphics[width=8cm,height=8cm,angle=0]{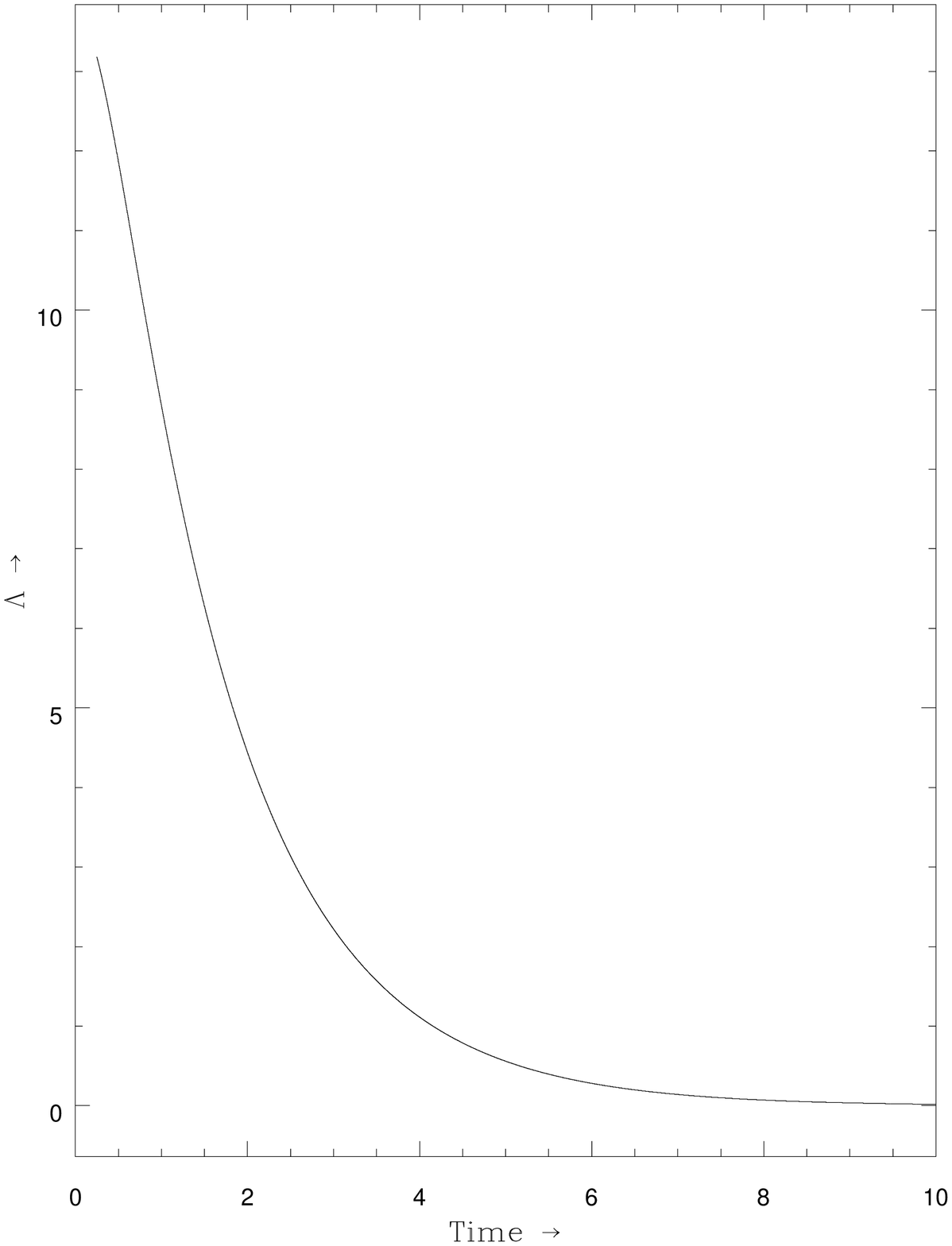}
\caption{The plot of cosmological term $\Lambda(T)$ Vs. T}
\end{figure}
\subsection{Model II: Solution when $m = 1$}
When $m = 1$, Eq. (\ref{eq35}) reduces to $\xi = \xi_{0}\rho$. With the use of Eqs. (\ref{eq33}), (\ref{eq34}) 
and (\ref{eq37}), Eq.  (\ref{eq36}) reduces to
\[
8\pi \rho = \frac{1}{a^{2}T_{1}^{\frac{2n(1 - \alpha)}{(1 + \beta)}}\left\{1 + \gamma - \frac{\xi_{0} K_{2}}
{T_{1}^{\frac{n(1 - \alpha)}{(1 + \beta)}}}\left(\frac{T_{2}}{T_{1}}\right)\right\}}\Biggl[\frac{b^{2}}{2(\beta + 1)^{2}}
\Big\{4n(1 - \alpha^{2}) + 2(\alpha + \beta) \,+  
\]
\begin{equation}
\label{eq41}
(\beta - \alpha)(1 - \alpha)(2\alpha + 1)\Big\}\left(\frac{T_{2}}{T_{1}}\right)^{2} - \, \frac{b^{2}(\alpha + 1)}
{(\beta + 1)} - 2K(1 + KX^{2})\Biggr].  
\end{equation}
Eliminating $\rho(t)$ between (\ref{eq33}) and (\ref{eq41}), we get
\[
\Lambda = \frac{1}{a^{2}T_{1}^{\frac{2n(1 - \alpha)}{(1 + \beta)}}\left\{1 + \gamma - \frac{\xi_{0} K_{2}}
{T_{1}^{\frac{n(1 - \alpha)}{(1 + \beta)}}}\left(\frac{T_{2}}{T_{1}}\right)\right\}}\Biggl[\frac{b^{2}}{2(\beta + 1)^{2}}
\Big\{4n(1 - \alpha^{2}) + 2(\alpha + \beta) \,+  
\]
\[
(\beta - \alpha)(1 - \alpha)(2\alpha + 1)\Big\}\left(\frac{T_{2}}{T_{1}}\right)^{2} - \, \frac{b^{2}(\alpha + 1)}
{(\beta + 1)} - 2K(1 + KX^{2})\Biggr].  
\]
\[
- \, \frac{1}{a^{2}T_{1}^{\frac{2n(1 - \alpha)}{(1 + \beta)}}}\Biggl[\frac{b^{2}}{2(\beta + 1)^{2}}
\Big\{2n(1 - \alpha^{2}) + 2\alpha +  
\]
\begin{equation}
\label{eq42}
(\beta - \alpha)(1 - \alpha)\Big\}\left(\frac{T_{2}}{T_{1}}\right)^{2} - \frac{b^{2}(1 - \alpha)}
{2(\beta + 1)} - K(2 + 3KX^{2})\Biggr].  
\end{equation}
\begin{figure}[htbp]
\centering
\includegraphics[width=8cm,height=8cm,angle=0]{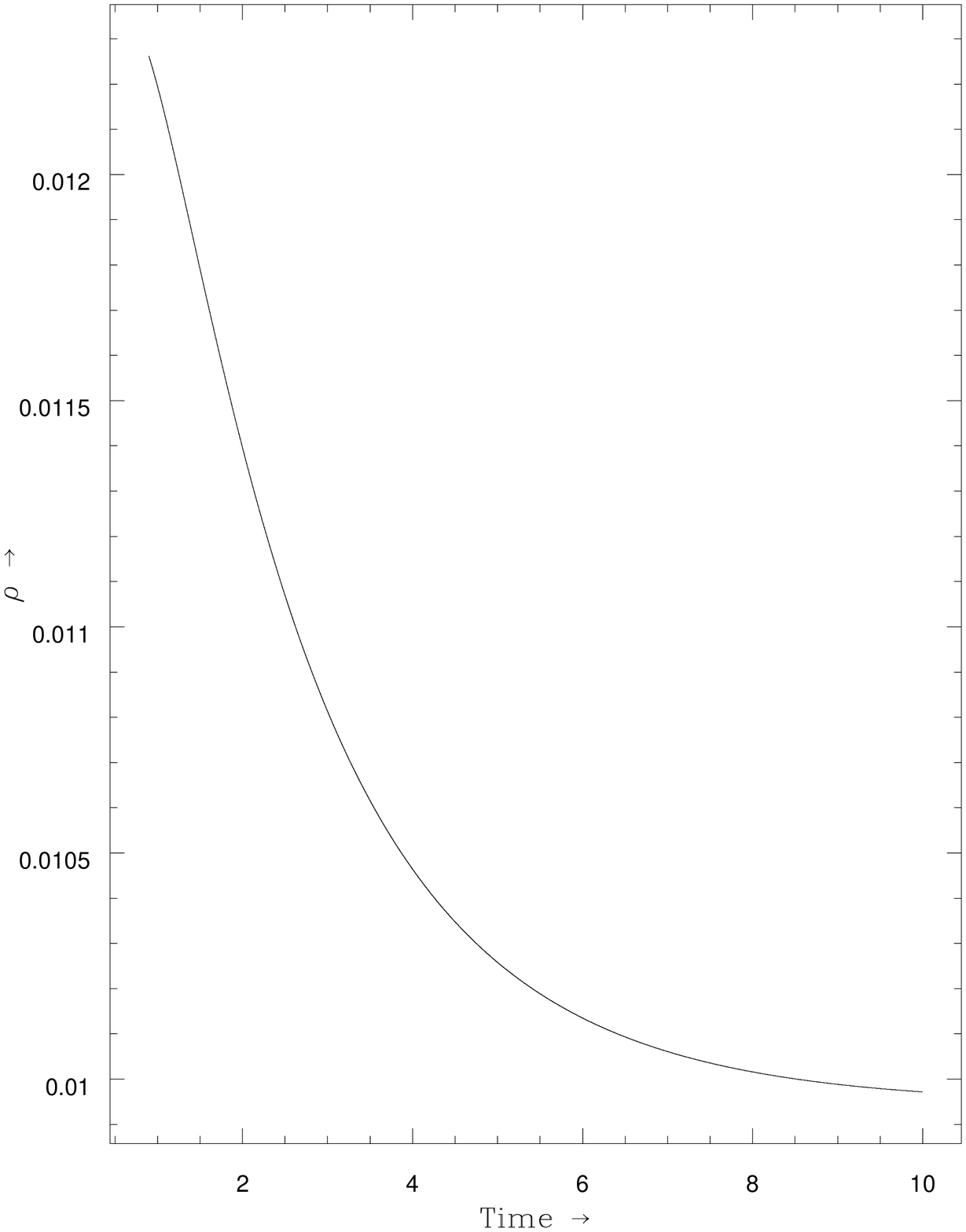}
\caption{The plot of energy density $\rho(T)$ Vs. T}
\end{figure}
\begin{figure}[htbp]
\centering
\includegraphics[width=8cm,height=8cm,angle=0]{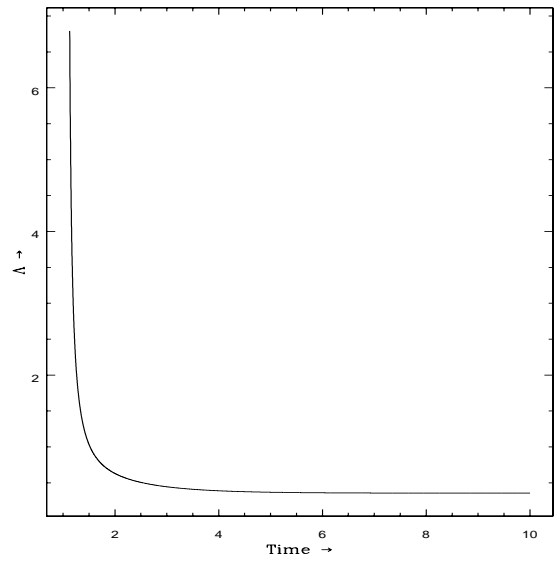}
\caption{The plot of cosmological term $\Lambda(T)$ Vs. T}
\end{figure}
From Eqs. (\ref{eq39}) and (\ref{eq41}) , we note that $\rho(t)$ is a decreasing function of time 
and $\rho > 0$ for all times. This behaviour is clearly depicted in Figures $1$ and $3$ as a representative 
case with appropriate choice of constants of integration and other physical parameters using reasonably 
well known situations. Figures $1$ and $3$ show this physical behaviours of energy density as a decreasing 
functions of coordinate time $T$ of Model I and II. Here the coordinate time $T$ is related to normal time as shown 
in Ref.\cite{ref19}. This also follows for rest part of the paper. \\ 

In spite of homogeneity at large scale our universe is inhomogeneous at small scales, so physical 
quantities being position dependent are more natural in our observable universe if we do not go 
to super high scale. This result shows this kind of physical importance. In recent time the $\Lambda$-term 
has interested theoreticians and observers for various reasons. The nontrivial role of the vacuum in 
the early universe generate a $\Lambda$-term that leads to inflationary phase. Observationally, this term 
provides an additional parameter to accommodate conflicting data on the values of the Hubble constant, the 
deceleration parameter, the density parameter and the age of the universe (for example, see the references 
\cite{ref75,ref76}). Assuming that $\Lambda$ owes its origin to vacuum interactions, as suggested in 
particular by Sakharov \cite{ref77}, it follows that it would in general be a function of space and time 
coordinates, rather than a strict constant. In a homogeneous universe $\Lambda$ will be at most time dependent 
\cite{ref78}. In our case this approach can generate $\Lambda$ that varies both with space and time. In 
considering the nature of local massive objects, however, the space dependence of $\Lambda$ cannot be ignored. 
For details discussion, the readers are advised to see the references (Narlikar, Pecker and Vigier 
\cite{ref79}, Ray and Ray \cite{ref80}, Tiwari, Ray and Bhadra \cite{ref81}).  \\ 

The behaviour of the universe in this model will be determined by the cosmological term $\Lambda$ ; this term 
has the same effect as a uniform mass density $\rho_{eff} = - \Lambda / 4\pi G$, which is constant in space 
and time. A positive value of $\Lambda$ corresponds to a negative effective mass density (repulsion). Hence, we 
expect that in the universe with a positive value of $\Lambda$, the expansion will 
tend to accelerate; whereas in the universe with negative value of $\Lambda$, the expansion will slow down, stop 
and reverse. From Eqs. (\ref{eq40}) and (\ref{eq42}), we see that the cosmological term $\Lambda$ is a decreasing 
function of time and it approaches a small positive value at late time. From Figures 2 and 4, we note this behaviour 
of cosmological term $\Lambda$ in both models I and II. Recent cosmological observations  suggest the existence 
of a positive cosmological constant $\Lambda$ with the magnitude $\Lambda(G\hbar/c^{3})\approx 10^{-123}$. These 
observations on magnitude and red-shift of type Ia supernova suggest that our universe may be an accelerating 
one with induced cosmological density through the cosmological $\Lambda$-term. Thus, our model is consistent 
with the results of recent observations. \\  

\noindent
{\bf{Some Physical and Geometric Features :}} \\
The non-vanishing component $F_{12}$ of electromagnetic field tensor is obtained as
\[
F^{2}_{12} = \frac{\bar{\mu}}{8\pi}\frac{b^{2}(1 - \alpha)}{(\beta + 1)}e^{KX^{2}}(c_{1}e^{bT} + 
c_{2}e^{-bT})^{\frac{2}{(\beta + 1)}}\times
\]
\begin{equation}
\label{eq43}
\Biggl[1 - \frac{(\beta - \alpha)}{(\beta + 1)}\frac{(c_{1}e^{bT} - c_{2}e^{-bT})^{2}}{(c_{1}e^{bT} + 
c_{2}e^{-bT})^{2}}\Biggr].
\end{equation}
The expressions for the shear scalar $\sigma^{2}$, acceleration vector $\dot{u}_{i}$ 
and proper volume $V^{3}$ for model (\ref{eq31}) are given by
\begin{equation}
\label{eq44}
\sigma^{2} = \frac{b^{2}\left[\{n(1 - \alpha) + (1 + \alpha)\}^{2} - 3n(1 - \alpha)(1 + \alpha) - 3\alpha\right]}
{3(\beta + 1)^{2}a^{2} (c_{1}e^{bT} + c_{2}e^{-bT})^{\frac{2n(1 - \alpha)}
{(\beta + 1)}}}\frac{(c_{1}e^{bT} - c_{2}e^{-bT})^{2}}{(c_{1}e^{bT} + 
c_{2}e^{-bT})^{2}},
\end{equation}
\begin{equation}
\label{eq45}
\dot{u}_{i} = (0, 0, 0, 0),
\end{equation}
\begin{equation}
\label{eq46}
V^{3} = \sqrt{-g} = a^{2}(c_{1}e^{bT} + c_{2}e^{-bT})^{\frac{2n(1 - \alpha) + (1 + \alpha)}{(\beta + 1)}},
\end{equation}
From Eqs. (\ref{eq44}) and (\ref{eq37}), we have
\begin{equation}
\label{eq47}
\frac{\sigma^{2}}{\theta^{2}} = \frac{\left[\{n(1 - \alpha) + (1 + \alpha)\}^{2} - 3n(1 - \alpha^{2}) 
-3\alpha\right]}{3\{n(1 - \alpha) + (1 + \alpha)\}^{2}} = \mbox{
constant}.
\end{equation}
The rotation $\omega$ is identically zero. From set of equations (\ref{eq37}), (\ref{eq43}) - (\ref{eq47}), 
the model brings out the following features: \\
The model starts expanding at $T>0$ and goes on expanding indefinitely when $\frac{n(1 - \alpha)}{(\beta + 1)} < 0$. 
The model (\ref{eq31}) represents an expanding, shearing and non-rotating universe in which the flow vector is geodetic. 
Since $\frac{\sigma}{\theta}$ = constant, the model does not approach isotropy. As $T$ increases 
the proper volume also increases. The model is non-accelerating. The physical quantities $p$ and $\rho$ 
decrease as $F_{12}$ increases. However, if $\frac{n(1 - \alpha)}{(\beta + 1)} > 0$, 
the process of contraction starts at $T>0$ and at $T = \infty$ the expansion stops. The electromagnetic 
field tensor does not vanish when $b \ne 0$, and $\alpha \ne 1$.
\section{Solution in Absence of Magnetic Field}
In absence of magnetic field the Einstein field equations for metric (\ref{eq1}) read as
\begin{equation}
\label{eq48}
\frac{1}{A^{2}}\left[- \frac{B_{44}}{B} - \frac{C_{44}}{C} + \frac{A_{4}}{A}\left(\frac{B_{4}}{B} + 
\frac{C_{4}}{C}\right) - \frac{B_{4} C_{4}}{B C} + \frac{B_{1}C_{1}}{BC}\right]  = 8 \pi p + \Lambda ,
\end{equation}
\begin{equation}
\label{eq49}
\frac{1}{A^{2}}\left(\frac{A^{2}_{4}}{A^{2}}- \frac{A_{44}}{A} - \frac{C_{44}}{C} + 
\frac{C_{11}}{C} \right) =  8 \pi p + \Lambda, 
\end{equation}
\begin{equation}
\label{eq50}
\frac{1}{A^{2}}\left(\frac{A^{2}_{4}}{A^{2}} - \frac{A_{44}}{A} - \frac{B_{44}}{B} + 
\frac{B_{11}}{ B}\right) =  8 \pi p + \Lambda, 
\end{equation}
\begin{equation}
\label{eq51}
\frac{1}{A^{2}}\left[- \frac{B_{11}}{B} - \frac{C_{11}}{C} + \frac{A_{4}}{A}\left(\frac{B_{4}}{B} + 
\frac{C_{4}}{C}\right) - \frac{B_{1}C_{1}}{BC}  + \frac{B_{4} C_{4}}{B C}\right] = 8 \pi \rho - \Lambda,
\end{equation}
\begin{equation}
\label{eq452}
\frac{B_{14}}{B} + \frac{C_{14}}{C} - \frac{A_{4}}{A}\left(\frac{B_{1}}{B} + \frac{C_{1}}{C}\right) = 0,
\end{equation}
Eqs. (\ref{eq49}) and (\ref{eq50}) lead to
\begin{equation}
\label{eq53}
\frac{B_{44}}{B} - \frac{B_{11}}{B} - \frac{C_{44}}{C} + \frac{C_{11}}{C} = 0.
\end{equation}
Eqs. (\ref{eq19}) and (\ref{eq53}) lead to
\begin{equation}
\label{eq54}
\frac{g_{44}}{g} - \frac{k_{44}}{k} = 0.
\end{equation}
Eqs. (\ref{eq23}) and (\ref{eq54}) lead to
\begin{equation}
\label{eq55}
\frac{g_{44}}{g} + \alpha \frac{g^{2}_{4}}{g^{2}} = 0,
\end{equation}
which on integration gives
\begin{equation}
\label{eq56}
g = (c_{4} t + c_{5})^{\frac{1}{(\alpha + 1)}},
\end{equation}
where $c_{4}$ and $c_{5}$ are constants of integration. Hence from (\ref{eq23}) and (\ref{eq56}), we have 
\begin{equation}
\label{eq57}
k = c(c_{4} t + c_{5})^{\frac{\alpha}{(\alpha + 1)}}.
\end{equation}
In this case (\ref{eq17}) leads to 
\begin{equation}
\label{eq58}
f = \exp{\left(\frac{1}{2}K(x + x_{0})^{2}\right)}.
\end{equation}
Therefore, we have
\begin{equation}
\label{eq59}
B = \exp{\left(\frac{1}{2}K(x + x_{0})^{2}\right)} (c_{4} t + c_{5})^{\frac{1}{(\alpha + 1)}},
\end{equation}
\begin{equation}
\label{eq60}
C = \exp{\left(\frac{1}{2}K(x + x_{0})^{2}\right)} c (c_{4} t + c_{5})^{\frac{\alpha}{(\alpha + 1)}},
\end{equation}
\begin{equation}
\label{eq61}
A = a (c_{4} t + c_{5})^{\frac{n(1 - \alpha)}{(1 + \alpha)}},
\end{equation}
where $a$ is already defined in previous section. \\
After using suitable transformation of the co-ordinates, the model (\ref{eq1}) reduces to the form
\[
ds^{2} = a^{2}(c_{4}T)^{\frac{2n(1 - \alpha)}{(1 + \alpha)}}(dX^{2} - dT^{2}) 
+ e^{KX^{2}}(c_{4} T)^{\frac{2}{(\alpha + 1)}} dY^{2}
\]
\begin{equation}
\label{eq62}
+ e^{KX^{2}}(c_{4} T)^{\frac{2\alpha}{(\alpha + 1)}} dZ^{2},
\end{equation}
where $x + x_{0} = X$, $y = Y$, $cz = Z$, $t + \frac{c_{5}}{c_{4}} = T$. \\

The expressions for effective pressure $\bar{p}$ and density $\rho$ for the model (\ref{eq62}) are given by 
\begin{equation}
\label{eq63}
8\pi \bar{p} = \frac{1}{a^{2}(c_{4}T)^{\frac{2n(1 - \alpha)}{(1 + \alpha)}}}\left[\frac{n(1 - \alpha^{2}) + \alpha}
{(\alpha + 1)^{2}}\frac{1}{T^{2}} + K^{2}X^{2}\right] - \Lambda,
\end{equation}
\begin{equation}
\label{eq64}
8\pi \rho = \frac{1}{a^{2}(c_{4}T)^{\frac{2n(1 - \alpha)}{(1 + \alpha)}}}\left[\frac{n(1 - \alpha^{2}) + \alpha}
{(\alpha + 1)^{2}}\frac{1}{T^{2}} - K(2 + 3KX^{2})\right] + \Lambda.
\end{equation}
On using (\ref{eq35}) in (\ref{eq63}), we obtain
\begin{equation}
\label{eq65}
8\pi (p - \xi_{0} \rho^{m} \theta)  = \frac{1}{a^{2}(c_{4}T)^{\frac{2n(1 - \alpha)}{(1 + \alpha)}}}
\left[\frac{n(1 - \alpha^{2}) + \alpha}{(\alpha + 1)^{2}}\frac{1}{T^{2}} + K^{2}X^{2}\right] - \Lambda,
\end{equation}
where $\theta$, in this case, is calculated for the flow vector $u^{i}$ and is given by
\begin{equation}
\label{eq66}
\theta = \frac{K_{3}}{T^{\frac{n(1 - \alpha) + (1 + \alpha)}{(1 + \alpha)}}},
\end{equation}
where
\begin{equation}
\label{eq67}
K_{3} =  \frac{n(1 - \alpha) + (1 + \alpha)}{a(1 + \alpha)c_{4}^{\frac{n(1 - \alpha)}{(1 + \alpha)}}}.
\end{equation}
\subsection{Model I: Solution when $m = 0$}
When $m = 0$, Eq. (\ref{eq35}) reduces to $\xi = \xi_{0}$. With the use of Eqs. (\ref{eq64}), (\ref{eq34}) 
and (\ref{eq66}), Eq.  (\ref{eq65}) reduces to
\begin{equation}
\label{eq68}
4\pi (1 + \gamma) \rho = \frac{1}{a^{2}(c_{4}T)^{\frac{2n(1 - \alpha)}{(1 + \alpha)}}}\left[\frac{n(1 - \alpha^{2}) + 
\alpha}{(\alpha + 1)^{2}}\frac{1}{T^{2}} - K(1 +  K X^{2})\right] +  \frac{4\pi \xi_{0}K_{3}}{T^{\frac{n(1 - \alpha) + 
(1 + \alpha)}{(1 + \alpha)}}}. 
\end{equation}
\begin{figure}[htbp]
\centering
\includegraphics[width=8cm,height=8cm,angle=0]{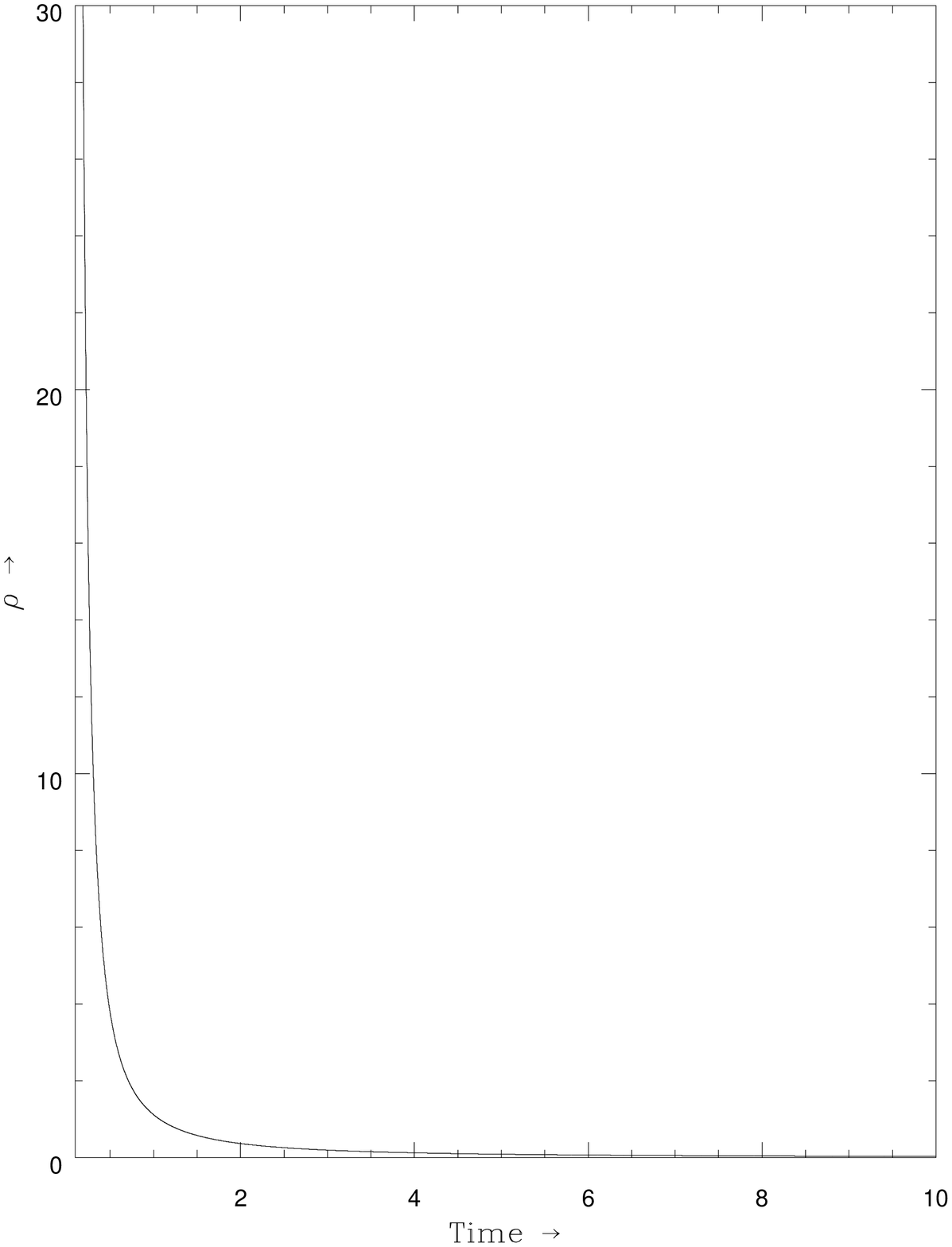}
\caption{The plot of energy density $\rho(T)$ Vs. T}
\end{figure}
\begin{figure}[htbp]
\centering
\includegraphics[width=8cm,height=8cm,angle=0]{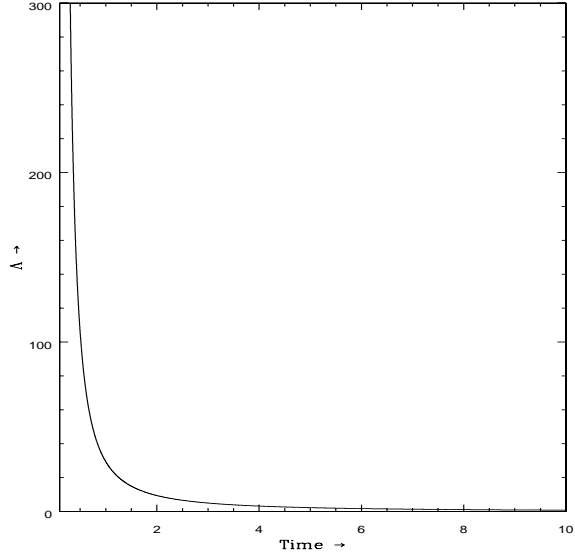}
\caption{The plot of cosmological term $\Lambda(T)$ Vs. T}
\end{figure}
Eliminating $\rho(t)$ between (\ref{eq64}) and (\ref{eq68}), we get
\[
(1 + \gamma)\Lambda = \frac{1}{a^{2}(c_{4}T)^{\frac{2n(1 - \alpha)}{(1 + \alpha)}}}\left[\frac{\{n(1 - \alpha^{2}) + 
\alpha\}}{(\alpha + 1)^{2}}\frac{(1 - \gamma)}{T^{2}} + K\{KX^{2}(1 +  3\gamma) + 2\gamma\}\right]
\]
\begin{equation}
\label{eq69}
+  \, \frac{8\pi \xi_{0}K_{3}}{T^{\frac{n(1 - \alpha) + (1 + \alpha)}{(1 + \alpha)}}}. 
\end{equation}
From Eq. (\ref{eq68}), we see that $\rho(t)$ is a decreasing function of time and $\rho > 0$ for 
all times. Figure $5$ shows this behaviour of energy density in Model I.\\ 
From Eq. (\ref{eq69}), we observe that the cosmological term $\Lambda$ is a decreasing function of time and it 
approaches a small positive value at late time. From Figure $6$, we note this behaviour 
of cosmological term $\Lambda$ in both model I. 
\subsection{Model II: Solution when $m = 1$}
When $m = 1$, Eq. (\ref{eq35}) reduces to $\xi = \xi_{0}\rho$. With the use of Eqs. (\ref{eq64}), (\ref{eq34}) 
and (\ref{eq66}), Eq.  (\ref{eq65}) reduces to
\[
4\pi \rho = \frac{1}{a^{2}(c_{4}T)^{\frac{2n(1 - \alpha)}{(1 + \alpha)}}\left\{1 + \gamma - \xi_{0}\frac{K_{3}}
{T^{\frac{n(1 - \alpha) + (1 + \alpha)}{(1 + \alpha)}}}\right\}}\times
\]
\begin{equation}
\label{eq70}
\left[\frac{n(1 - \alpha^{2}) + \alpha}{(\alpha + 1)^{2}}\frac{1}{T^{2}} - K(1 +  K X^{2})\right]. 
\end{equation}
\begin{figure}[h] 
\vspace*{13pt}
\includegraphics[width=11cm,height=11cm,angle=0]{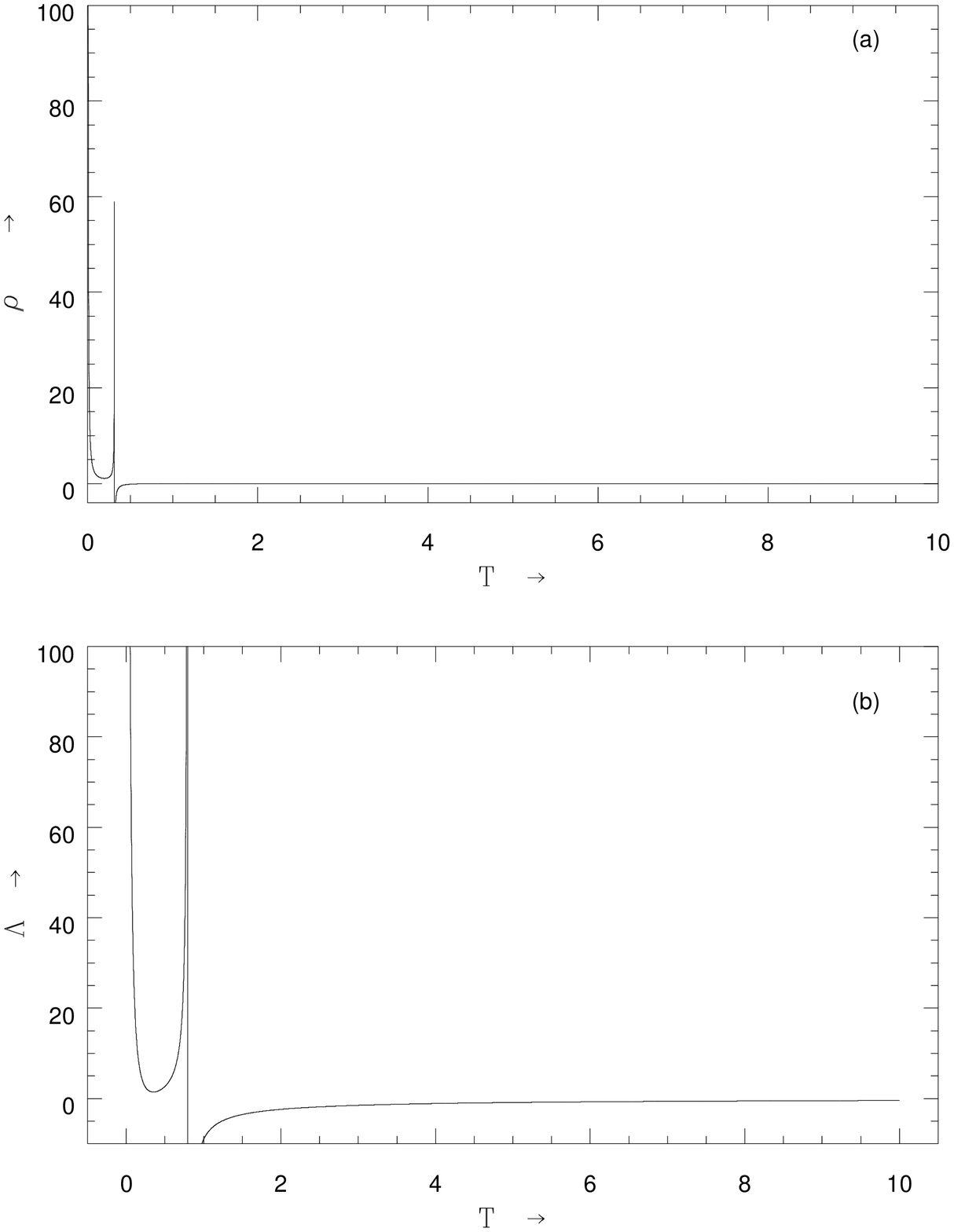}
\vspace*{13pt}
\caption{The plot of (a) $\rho$ as a function of T, 
(b) $\Lambda$ as a function of T}
\label{fig1}
\end{figure}
Eliminating $\rho(t)$ between (\ref{eq64}) and (\ref{eq70}), we get
\[
\Lambda = \frac{2}{a^{2}(c_{4}T)^{\frac{2n(1 - \alpha)}{(1 + \alpha)}}\left\{1 + \gamma - \xi_{0}\frac{K_{3}}
{T^{\frac{n(1 - \alpha) + (1 + \alpha)}{(1 + \alpha)}}}\right\}}\times
\]
\[
\left[\frac{n(1 - \alpha^{2}) + \alpha}{(\alpha + 1)^{2}}\frac{1}{T^{2}} - K(1 +  K X^{2})\right]
\]
\begin{equation}
\label{eq71}
- \, \frac{1}{a^{2}(c_{4}T)^{\frac{2n(1 - \alpha)}{(1 + \alpha)}}}\left[\frac{n(1 - \alpha^{2}) + \alpha}
{(\alpha + 1)^{2}}\frac{1}{T^{2}} - K(2 + 3KX^{2})\right]
\end{equation}
From Eq. (\ref{eq70}) and also from Figure $7(a)$, it seems that in very early stage of universe the energy density 
could be negative and may be link with some of the early universe physics which is the domain of quantum cosmology 
or early quantum mechanics. So we do not infer any things about energy density. But note that once energy density 
is negative in the initial stage even after oscillations it returns back to small negative value at the later stage 
of the evolution. Hence the details of macro physics in early universe decides its fate. So we cannot  infer from 
this model. Similar behaviours is also reflected about the cosmological constant $\Lambda$. We do not know for 
wide range of parameters this behaviour holds. It may be feasible that Figures $7(a, b)$ may coincide with Figures 
$3$ and $4$. But it does not seem to be vary generic behaviour of generalized model of section $3.2$. So we do not 
go to any conclusion about its applicability in general. But it seems that magnetic field and bulk viscosity prevent 
this kind of unusual behaviours as it is clear from Figures $1$ to $4$. \\   

\noindent
{\bf{Some Physical and Geometric Features}}: \\

The expressions for the shear scalar $\sigma^{2}$, acceleration vector $\dot{u}_{i}$ 
and proper volume $V^{3}$ for model (\ref{eq62}) are given by
\begin{equation}
\label{eq72}
\sigma^{2} = \frac{\{n(1 - \alpha) + (1 + \alpha)\}^{2} - 3n(1 - \alpha^{2}) - 3\alpha}{3(1 + \alpha)^{2}
a^{2}c_{4}^{\frac{2n(1 - \alpha)}{(1 + \alpha)}} T^{\frac{2n(1 - \alpha) + 2(1 + \alpha)}{(1 + \alpha)}}},
\end{equation}
\begin{equation}
\label{eq73}
\dot{u}_{i} = (0, 0, 0, 0),
\end{equation}
\begin{equation}
\label{eq74}
V^{3} = \sqrt{-g} = a^{2}e^{KX^{2}}(c_{4}T)^{\frac{2n(1 - \alpha) + (1 + \alpha)}{(1 + \alpha)}}.
\end{equation}
From (\ref{eq72}) and (\ref{eq66}), we obtain
\begin{equation}
\label{eq75}
\frac{\sigma^{2}}{\theta^{2}} = \frac{\{n(1 - \alpha) + (1 + \alpha)\}^{2} - 3n(1 - \alpha^{2}) - 3\alpha}
{3\{n(1 - \alpha) + (1 + \alpha)\}^{2}}.
\end{equation}
The rotation $\omega$ is identically zero.\\

The model in absence of magnetic field starts expanding with a big bang at $T = 0$ and it stops 
expanding at $T = \infty$. In absence of magnetic field, the model in general represents an 
expanding, shearing and non-rotating in which the flow vector is geodetic. Since $\frac{\sigma}{\theta}$ 
= constant, the model does not approach isotropy. As $T$ increases the proper volume also increases. 
The model is non-accelerating. 
\section{Concluding Remarks}
We have obtained a new cylindrically symmetric inhomogeneous cosmological model of electro-magnetic
bulk viscous fluid as the source of matter. Generally the model represents expanding, shearing and non-
rotating universe in which the flow vector is geodetic. It is worth mention here that in presence 
of magnetic field the model (\ref{eq31}) is expanding whereas in absence of magnetic field the model 
(\ref{eq62}) starts with a big bang singularity. In both models $\frac{\sigma}{\theta} =$ constant 
and hence they do not approach isotropy. The models are non-accelerating. In these solutions all 
physical quantities depend on at most one space co-ordinate and time.\\
It is important to note here that both the models (\ref{eq31}) and (\ref{eq62}) in presence and 
absence of magnetic field reduce to homogeneous universe when $K = 0$. This shows that 
for $K = 0$, inhomogeneity dies out. \\

The effect of bulk viscosity is to produce a change in perfect fluid and hence exhibit essential influence 
on the character of the solution. The effect is clearly visible on the $p$ effective (see details in previous 
sections). In Section $3$, we have shown regular well behaviour of energy density, cosmological constant ($\Lambda$) 
and the expansion of the universe with parameter $T$.  The section $4$ is a toy investigation to see that the 
effect of bulk viscosity plays dynamic role in the evolution equations. We also observe that Murphy's conclusion 
\cite{ref65} about the absence of a big bang type singularity in the infinite past in models with bulk viscous 
fluid, in general, is not true. The results obtained by Myung and Cho \cite{ref60} also show that, it is, in general, 
not valid, since for some cases big bang singularity occurs in finite past. \\

In presence and absence of magnetic field, the cosmological terms in models are decreasing function of time 
and approach a small value at late time (with exception $m = 1$ in the absence of magnetic field). The values 
of cosmological ``constant'' for the models are found to be small and positive, as obtained in  recent 
results from the supernovae observations recently obtained by the High-Z supernovae Ia Team and Supernovae 
Cosmological Project. Our solutions generalize the solutions recently obtained by Pradhan et al. \cite{ref66}.    
\section*{Acknowledgments} 
Authors would like to thank the Inter-University Centre for Astronomy and Astrophysics (IUCAA), Pune, 
India for providing facility and support where this work was carried out. A. P. and K. J. are 
visiting associates of IUCAA.   

\end{document}